\documentclass[lettersize,journal,twocolumn]{IEEEtran}
\usepackage{amsmath,amsfonts}
\usepackage{algorithmic}
\usepackage{algorithm}
\usepackage{array}
\usepackage{textcomp}
\usepackage{stfloats}
\usepackage{xcolor}
\usepackage{url}
\usepackage{verbatim}
\usepackage{graphicx}
\usepackage{tabularx,ragged2e,booktabs}
\usepackage[compatibility=false]{caption}
\usepackage{nicematrix}
\usepackage[greek,english]{babel}
\newtheorem{Theorem}{Theorem}

\begin{document}
	\title{A Mathematical Model of Hybrid Microgrid With Pole Placement Controller Using State Feedback For Stability Improvement}

	\author{Yangyadatta~Tripathy,~\IEEEmembership{Graduate Member,~IEEE,}
        Barjeev~Tyagi,~\IEEEmembership{Member,~IEEE,}}
	\maketitle

	\begin{abstract}
		This paper presents the development of a mathematical model of a converter state space model for a hybrid microgrid. The hybrid model combines the models of components such as DC-Converters, DC-AC converters, and their individual controllers, as well as loads. The input to the converter is considered a constant DC voltage, assumed to originate from distributed generations like solar, battery storage, or fuel-cells. The converter output is connected to a DC line through an LCL filter. The controller circuitry is designed to regulate the voltage, current, and power from the converter. Sensors are strategically placed to measure the currents, voltages, and power, and calculate the reference pulse signal using PWM for the switch. Similarly, the DC-AC converter is modeled. In the state space domain the converter models is used to design overall microgrid system. A single DC converter has six states and two inputs, with all states as outputs. A single DC-AC converter has thirteen states and three inputs, with all states as outputs. Three such converters of each type are considered to develop the DC microgrid and AC microgrid, which are then combined using mathematical analysis to model a hybrid microgrid. For the hybrid microgrid development, network and load models were also included. Eigenvalue analysis has been conducted to study the small signal stability of the considered system. The complete state space model of the hybrid microgrid has been programmed, and a pole-placement controller has been designed to enhance the stability of the system.
	\end{abstract}

		
\maketitle
\section{Introduction}
The usage of renewable energy sources is being popular due to low carbon footprint and to minimize environmental polution \cite{Tripathy2023} resulting to slowdown climate change. The integration of the energy sources like solar and wind energy etc. with the conventional energy sources is required. Microgrid infrastructure has been a promising solution to augment conventional plant stability and offering efficient energy distribution. Microgrids offer several advantages, such as higher efficiency \cite{Drag2016}, natural integration of renewable-based generation and storage devices, and compatibility with consumer electronics. Renewable-based generations coupled with storage devices in microgrids \cite{Demello1969} can be made dispatchable, which is an essential feature of this technology. With the rise of electronic components in domestic loads, the compatibility with consumer electronics is also another benefit \cite{Drag2016}. As a result, microgrids have gained popularity and are often considered for various applications, including community-based systems in remote and developing regions. In microgrids, storage systems are employed as dispatchable units alongside renewable generations to enable voltage regulation and stability retention. However, the widespread adoption of storage systems is often hindered by their high capital cost, which remains a significant barrier. Nevertheless, the integration of storage systems with renewable-based generation in microgrids remains a crucial strategy for enhancing the efficiency, reliability, and flexibility of power distribution systems. Storage systems are used as dispatchable units alongside renewable generation for applications like voltage regulation and stability retention in microgrids. However, the primary barrier to the widespread use of storage for these applications is its high capital cost \cite{Gui2021a}.

As microgrids become more widely deployed, controllers are essential for maintaining system stability. Integrating controllers ensures the system states remain within acceptable bounds, improving stability in complex dynamic situations. Several research papers have reported stability studies  \cite{Wang2018}, \cite{Rashidirad2018} on DC microgrid operations, discussing small-signal stability, also known as dynamic or steady-state stability. These studies observe the behavior of system states in response to small perturbations to validate small-signal stability. For broader analyses, such as transient or fault analysis, large-signal studies are conducted. Microgrids are susceptible to instability when subjected to constant power loads (CPL) due to the negative impedance characteristics of these loads. In \cite{Wang2018}, the voltage stability issue is addressed by modeling the system’s stability as a parallel impedance curve. The presence of CPL introduces unstable poles, which degrade the system's stability. To mitigate this instability, methods such as filtering, load shedding, or appropriate control strategies are required. The parallel operation of microgrids necessitates droop control, which can lead to low-frequency instabilities that demand prompt attention. In \cite{Rashidirad2018}, the correction of low-frequency oscillations in DCMGs is explored. Virtual impedance is employed in \cite{Lu2015a} to enhance stability for constant power loads. CPL is identified as the primary cause of dynamic interactions in DCMGs, with distributed generators, load dynamics, and various resonant phenomena influencing high-frequency voltage and current oscillations. A detailed analysis of high-frequency dominant modes and their damping is presented in \cite{Lu2015a}. Furthermore, an eigenvalue-based study on the small-signal model of a community DC microgrid that integrates distributed DC electric springs is conducted [8]. DC electric springs, which are demand-side management technologies, help reduce storage needs in a microgrid. This study evaluates the destabilizing effects of CPLs and performs a sensitivity analysis of frequency modes in response to control parameter variations. 

For large load variations or faults in a microgrid, large signal analysis becomes essential. This approach uses nonlinear mathematical methods to assess the system's behavior under significant disturbances. Large signal analysis covers a broader domain than small signal analysis and depends on the system's order. In \cite{Microgrids2021, Kabalan2017, Dongdong2019}, large signal analysis is applied, and islanded DCMG modeling is performed. For various load types, including constant impedance, constant current, and constant power in parallel boost-converter-based DC microgrids, a current-limiting droop controller is designed in \cite{Braitor2021}. This controller ensures closed-loop stability and power sharing. However, the main limitation of current-limiting controllers is their ineffectiveness during transient conditions, as they are saturation-based. This leads to issues like integrator windup, where the integral control action exceeds the saturation limits of the control element, preventing reliable closed-loop stability. In research papers, the small-signal analysis of microgrids, around the operating points, the low-frequency oscillations are present. To address these issues of low-frequency instabilities, a robust state feedback controller is required to integrate in the proposed model \cite{Kautsky1985}. The state feedback controller plays an important role for placing the poles of unstable system in desired locations within 10\% accuracy which makes the system Lyapunov stable. The proposed model ensures closed-loop stability during load-transients. 

    The aim and contribution of the paper are as follows.
	\begin{enumerate}
			\item A DC microgrid state space model is developed considering the dynamics of DC-DC converter and filter and connector as power circuit and current and voltage controllers.
			\item An AC microgrid model is developed considering the dynamics of ac converters and filters as well as bus-connecting lines and loads. An interlinking converter was modeled implementing swing equation and power controller for bidirectional power flow between ac and dc microgrids.
			\item A State feedback pole-placement controller was implemented to make unstable poles of the modelled system stay in the desired locations for stability. Stability region with respect to droop control gains were established.
	\end{enumerate}
	
	\subsection{Organisation of the paper} This paper is organized as follows. In section II, the small signal model for a complete DC Microgrid components i.e. DC-DC converter is developed, representing the dynamics of individual current, voltage, and power controller in terms of differential equations. In section III the small signal model for a complete AC Microgrid components i.e. DC-AC converter is developed, representing the dynamics of individual current, voltage controller, and in section IV the hybrid test system is demonstrated using DC sub-grid and AC sub-grid. In section V, the pole-placement technique using state feedback is explained. Then in section VI, a stability study is conducted using eigenvalue locations and pole-zero locations. To understand the system dynamics better, step responses of the states were observed with and without state feedback control. Section VI concludes the paper.
	\section{DC-DC Converter System Modeling for DC Sub-grid}
\begin{figure}[H]
		\centering
		\includegraphics[width=0.9\linewidth]{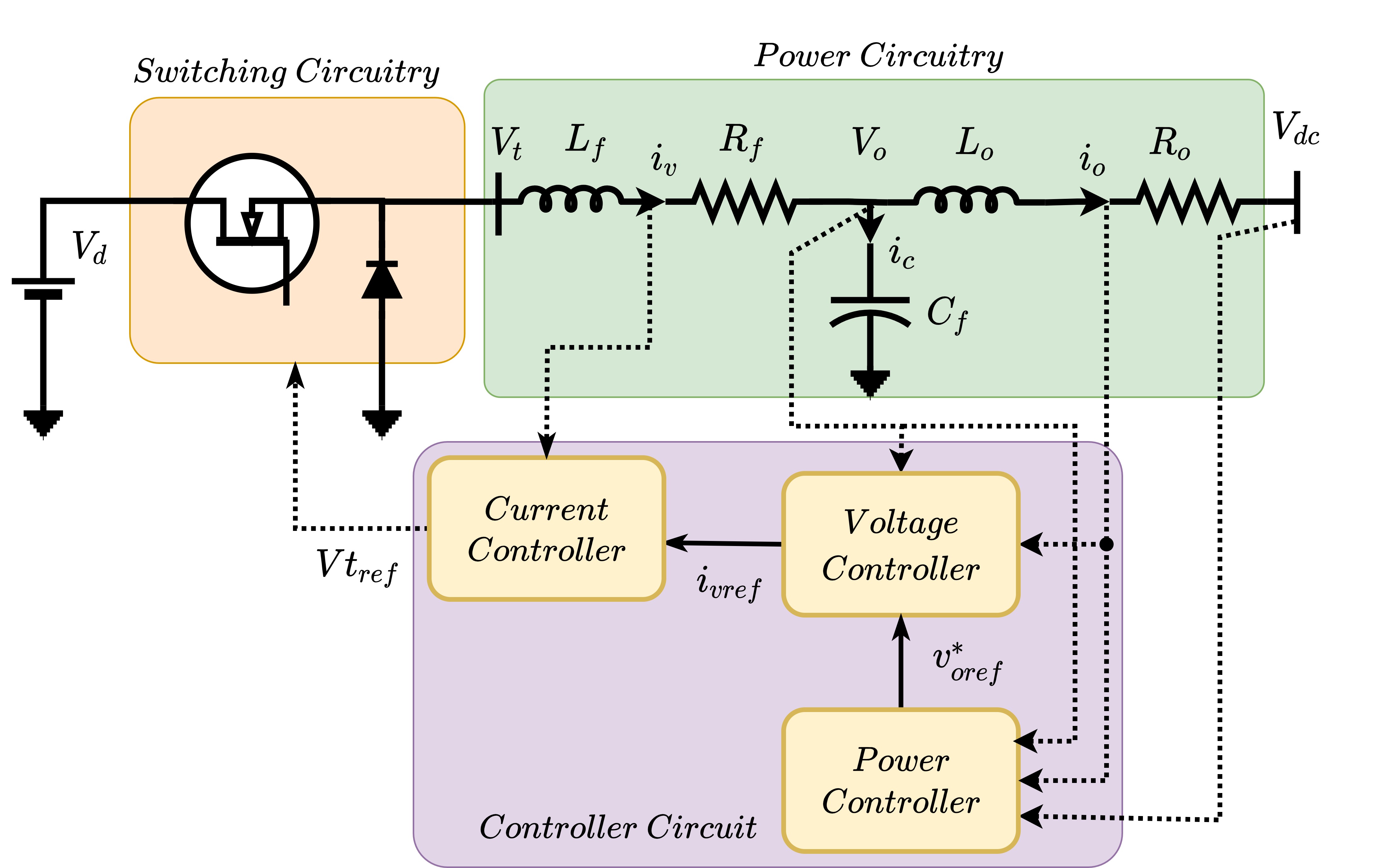}
		\caption{Block diagram representation of the DC-DC Converter}
		\label{fig:convertermodel}
	\end{figure}

	In this paper, a DC-DC step-down converter-based DC microgrid is developed to represent the dynamic behavior of a DC microgrid system. The converter comprises three distinct circuit subsystems: the switching circuitry, the power circuitry, and the control circuitry, as illustrated in Fig. \ref{fig:convertermodel}. The switching circuit includes key components such as a switch and a power diode, which regulate the flow of current. The power circuitry incorporates an LC filter, which smooths the output voltage, and a connector that is modeled as an RL element to capture the behavior of the connecting components. The control circuitry consists of three fundamental controllers—voltage, current, and power controllers—that manage the operation of the converter and ensure stable performance. 
    \par The dynamic models of these individual control components are formulated through the use of differential equations, which describe the time-varying behavior of the system. To facilitate further analysis, these models are subsequently represented in state-space form, providing a comprehensive mathematical framework for studying the system’s stability and performance. The state-space model is derived and discussed in the following sections, offering a detailed understanding of the interactions between the control components and the power electronics within the DC microgrid.*
	\subsection{Modelling of power circuit of the DC-DC Converter}
	The differential equation governing the power circuit of the DC-DC converter can be expressed by the following equation. By applying Kirchhoff's Voltage Law (KVL) to the filter and output circuits, along with Kirchhoff's Current Law (KCL) at the 
$V_0$ node, the resulting equations are derived. These equations represent the dynamic behavior of the converter's components, including the inductor, capacitor, and the control signals. The system's response is then analyzed to ensure stability and desired performance characteristics under various operating conditions.
	\begin{equation}
		\frac{\partial i_v}{\partial t} = \frac{1}{L_f}(V_t-V_o) - \frac{R_f*i_v}{L_f}
	\end{equation}
	\begin{equation}
		\frac{\partial i_o}{\partial t} = \frac{1}{L_o}(V_o-V_{dc}) - \frac{R_o*i_o}{L_o}
	\end{equation}
	\begin{equation}
		\frac{\partial v_o}{\partial t} = \frac{1}{C_f}(i_v-i_{o})
	\end{equation}
	where $R_f$, $L_f$ and $C_f$ are resistance, inductance and capacitance of the filter respectively and $R_o$, $L_o$ denotes the resistance and inductance parameter of the output connector which acts as the interface between the converter and DC bus. $V_t$ is the terminal voltage of the converter, $V_o$ is the capacitor voltage and $V_{dc}$ corresponds to the DC bus voltage.
	The state space model is expressed as following.
	\begin{multline*}
		\begin{bmatrix}
			\Delta\dot{i_v} \\ 	\Delta\dot{i_o} \\ 	\Delta\dot{V_o}
		\end{bmatrix} = 
		\begin{bmatrix}
			\frac{-R_f}{L_f} & 0 &\frac{1}{L_f}\\ 0 & \frac{-R_o}{L_o} & \frac{1}{L_o} \\ \frac{1}{C_f} & \frac{-1}{C_f} & 0
		\end{bmatrix}
		\begin{bmatrix}
			\Delta{i_v} \\ \Delta{i_o} \\ \Delta{V_o}
		\end{bmatrix} +
		\begin{bmatrix}
			\frac{1}{L_f} \\ 0 \\ 0
		\end{bmatrix} 
		\begin{bmatrix}
			\Delta{V_t}
		\end{bmatrix} + \\
		\begin{bmatrix}
			0 \\ \frac{-1}{L_o} \\ 0
		\end{bmatrix} 
		\begin{bmatrix}
			\Delta{V_{dc}}
		\end{bmatrix}
	\end{multline*}
	The complete state model for the power circuit of the DC-DC converter is described as:
	\begin{equation}\label{eqn4}
		\big[\Delta\dot{X_p}\big] = \big[A_p\big]\big[\Delta X_p\big]+\big[B_{1p}\big]\big[\Delta V_t\big]+\big[B_{2p}\big]\big[\Delta V_{dc} \big]
	\end{equation}
	The $\Delta$ before the quantities expresses small change in parameter representing small signal modelling. $\dot{Q}$ represents the derivative of $Q$ with respect to time.
	\subsection{Modelling of voltage controller of the DC-DC converter}
\begin{figure}[H]
		\centering
		\includegraphics[width=0.9\linewidth]{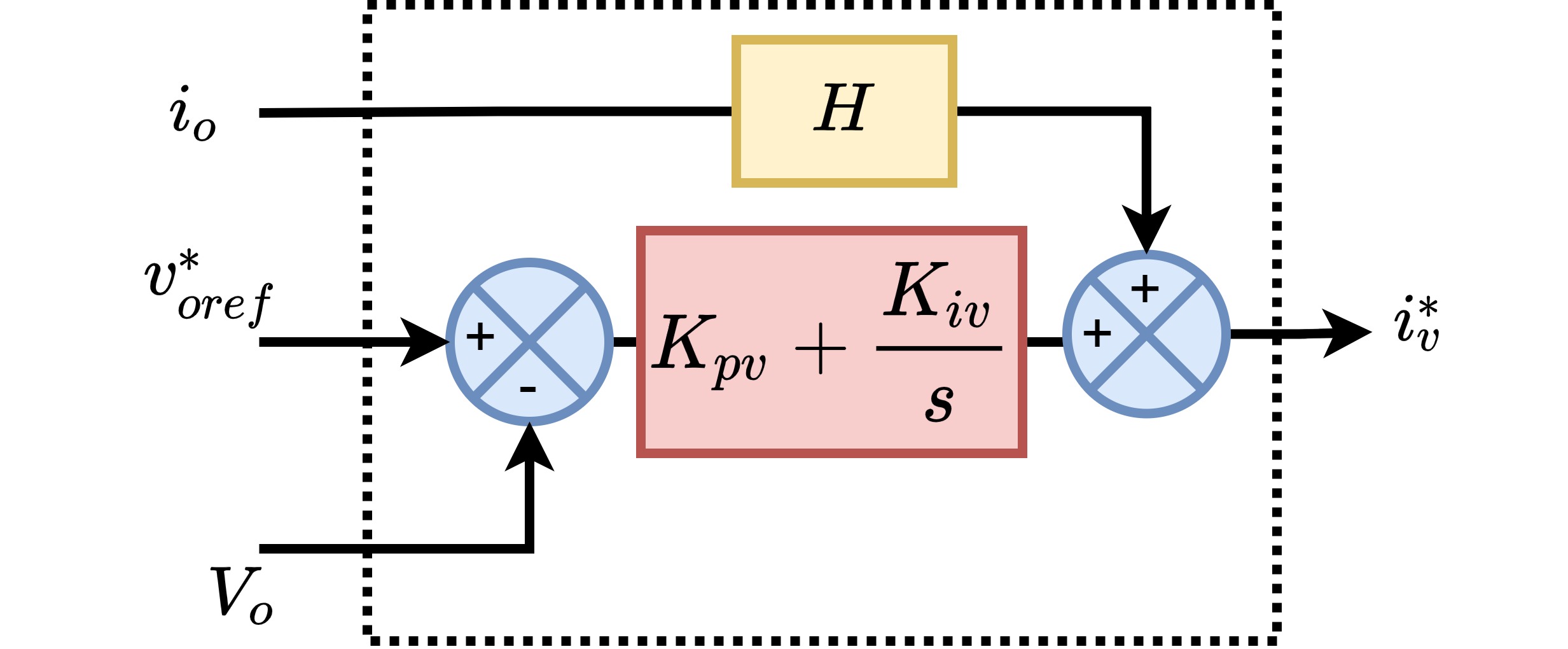}
		\caption{Block diagram representation of the voltage controller for DC-DC Converter}
		\label{fig:voltagecontroller}
	\end{figure}

	The voltage controller is composed of a PI controller and a feedforward gain applied to the output current of the DC-DC converter, which establishes the reference input for the current controller, as depicted in Fig. \ref{fig:voltagecontroller}. The current control loop functions as an inner current loop, as illustrated in Fig. \ref{fig:convertermodel}. To enhance control accuracy, a knowledge-based feedforward gain is utilized for the output current.
	The PI controller processes the error between the commanded voltage and measured voltage as depicted.
	\vspace{-10pt}	
	\begin{equation*}
		i_v^* = \big[V_o-V_{dc}\big]\big[K_{pv}\big]+\big[V_o-V_{dc}\big]\big[\frac{K_{iv}}{s}\big]+\big[H*i_o\big]
	\end{equation*}
	where $K_{pv}$ and $K_{iv}$ are proportional and integral gains of the PI controller used. A state variable $\zeta_{v}$ is introduced, which takes the integral output of error between commanded and measured voltage.
	\begin{equation*}
		\zeta_v = \big[V_o-V_{dc}\big]\big[\frac{K_{iv}}{s}\big]
	\end{equation*}
	The state space model for the voltage controller can be described using the following equations:
	\begin{equation}\label{eq5}
		\big[\Delta\dot{i_v^*}\big] = \big[1\big]\big[\Delta \zeta_{v}\big]+\big[K_{pv}\big]\big[\Delta V_o\big]+\big[-K_{pv}\big]\big[\Delta V_{dc}\big]+\big[H\big]*\big[i_o\big]
	\end{equation}
	\begin{equation}\label{eq6}
		\big[\Delta\dot{\zeta_v}\big] = \big[0\big]\big[\Delta \zeta_{v}\big]+\big[K_{iv}\big]\big[\Delta V_o\big]+\big[-K_{iv}\big]\big[\Delta V_{dc}\big]
	\end{equation}	
	\subsection{Modelling of current controller of the DC-DC converter}
\begin{figure}[H]
		\centering
		\includegraphics[width=0.9\linewidth]{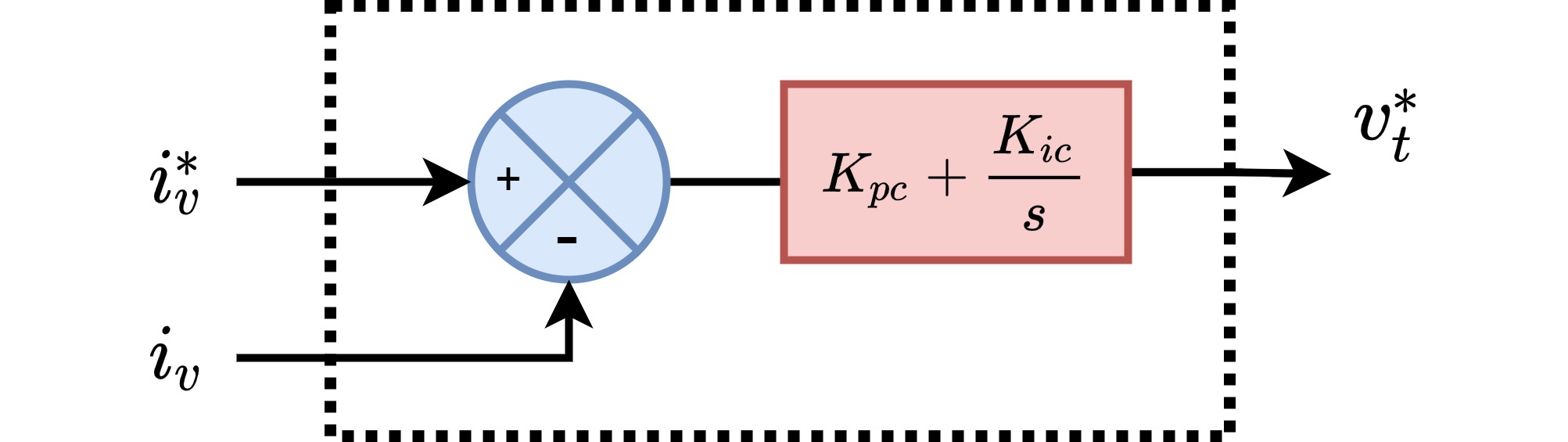}
		\caption{Block diagram representation of the current controller for DC-DC Converter}
		\label{fig:currentcontroller}
	\end{figure}

	The current controller is implemented using a PI controller, which calculates the error between the reference output current and the actual output current of the DC-DC converter. This error is subsequently used to determine the reference output voltage for the converter, as illustrated in Fig. \ref{fig:currentcontroller}.
	The small signal model of the current controller can be explained using the below differential equation.
	\begin{equation*}
		V_t = \big[i_v^*-i_{v}\big]\big[K_{pc}\big]+\big[i_v^*-i_{v}\big]\big[\frac{K_{ic}}{s}\big]
	\end{equation*}
	where $K_{pc}$ and $K_{ic}$ are proportional and integral gain of the PI controller for current control. A state variable $\eta_{c}$ is introduced which takes the integral output of error between reference and measured current.
	\begin{equation*}
		\eta_c = \big[i_v^*-i_{v}\big]\big[\frac{K_{ic}}{s}\big]
	\end{equation*}
	The state space model for the current controller can be described using the following equations:
	\begin{equation}\label{eq7}
		\big[\Delta\dot{V_t}\big] = \big[1\big]\big[\Delta \eta_{c}\big]+\big[K_{pc}\big]\big[\Delta i_v^*\big]+\big[-K_{pc}\big]\big[\Delta i_{v}\big]
	\end{equation}
	\begin{equation}\label{eq8}
		\big[\Delta\dot{\eta_c}\big] = \big[0\big]\big[\Delta \eta_{c}\big]+\big[K_{ic}\big]\big[\Delta i_v^*\big]+\big[-K_{ic}\big]\big[\Delta i_{v}\big]
	\end{equation}
	\subsection{Modelling of DC Power controller of the DC-DC converter}
	The power controller for the DC-DC converter consists of a low pass filter with cut-off frequency $\omega_f$ and a voltage-power droop
	as described below. A alternate to voltage-power droop is voltage-current droop where instead of droop gain the virtual resistance is considered \cite{Drag2016}.
	\vspace{-12pt}	
	\begin{figure}[H]
		\centering
		\includegraphics[width=0.9\linewidth]{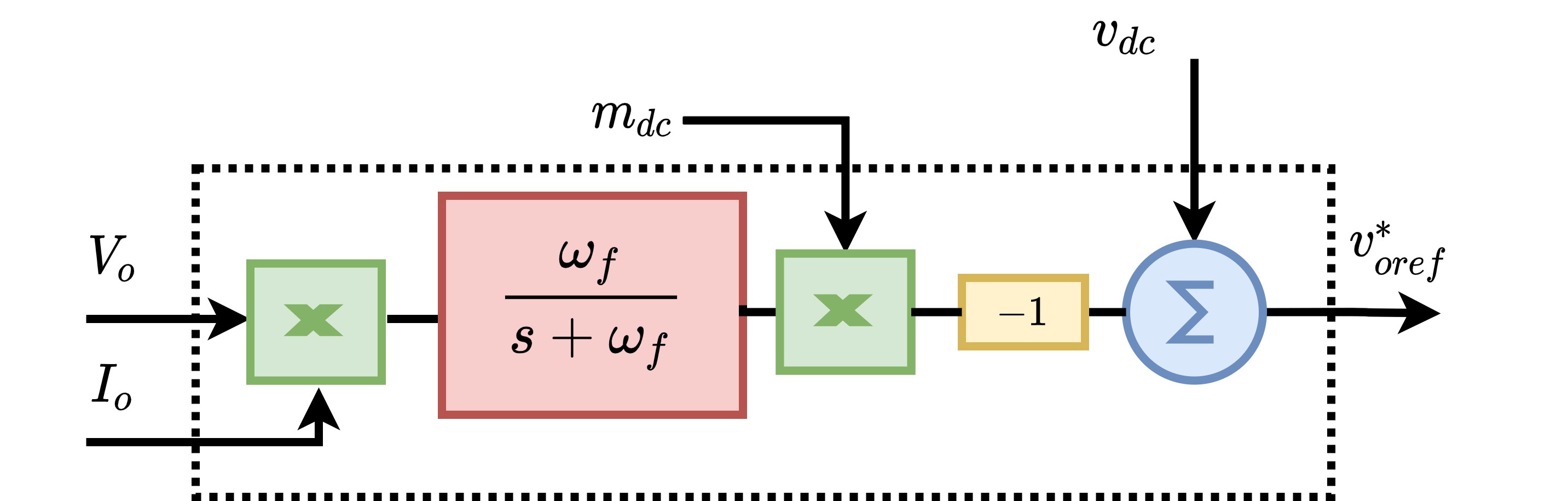}
		\caption{Block diagram representation of the DC power controller for DC-DC Converter}
		\label{fig:powercontroller}
	\end{figure}
	The DC power is filtered with a low pass filter with cut off frequency $\omega_f$ which imposes non-linearity due to the product term of voltage and current as both are the states of the converter.
	\begin{equation*}
		P_{dc} = \frac{\omega_f}{s+\omega_f}(i_o*v_o)
	\end{equation*}
	The above equation is linearized as described in the eqn \ref{eq9}:
	\begin{equation}\label{eqn9}
		\big[\Delta\dot{P_{dc}}\big] = \big[\omega\cdot v_o\big]\big[\Delta i_o\big]+\big[\omega\cdot i_o\big]\big[\Delta v_o\big]-\big[\omega\big]\big[\Delta P_{dc}\big]
	\end{equation}
	The voltage-power droop equation is presented as follows. For this $\Delta P_{dc}$ is the filtered power
	\begin{equation}\label{eq10}
		v_{oref}^* = v_{dc} - m_{dc}(\Delta P_{dc})
	\end{equation}
	where  $m_{dc}$ is the DC droop gain.
	The overall state space model for the power controller is described as follows.
	\begin{equation}\label{eq11}
		\big[\Delta{V_{o}}\big] = \big[- m_{dc}\big]\big[\Delta P_{dc}\big]
	\end{equation}
	using eqn \eqref{eq11} in eqn \eqref{eqn5} 
	\begin{multline}\label{eq12}
		\big[\Delta\dot{i_v^*}\big] = \big[1\big]\big[\Delta \zeta_{v}\big]+\big[K_{pv}\big]\big[m_{dc}\big]\big[\Delta P_{dc}\big] \\
		+\big[-K_{pv}\big]\big[\Delta V_{dc}\big]+ \big[H\big]*\big[i_o\big]
	\end{multline}
	using eqn \eqref{eq12} in eqn \eqref{eq7} the final state $V_t$ for the power controller is as follows.
	\begin{multline}\label{eq13}
		\big[\Delta\dot{V_t}\big] = \big[1\big]\big[\Delta \eta_{c}\big]+\big[K_{pc}\big]\bigg[\big[1\big]\big[\Delta \zeta_{v}\big]+\big[K_{pv}\big]\big[m_{dc}\big]\big[\Delta P_{dc}\big] \\
		+\big[-K_{pv}\big]\big[\Delta V_{dc}\big]+\big[H\big]*\big[i_o\big]\bigg]+\big[-K_{pc}\big]\big[\Delta i_{v}\big]
	\end{multline}
	%---------------------------------------------------------------------------------------------------------------------------------------------
	%---------------------------------------------------------------------------------------------------------------------------------------------
	\subsection{State Space model of the whole DC-DC Converter along with controllers}
	The DC-DC converter modeled in this paper comprises a voltage, current, and power controller for parameter regulation, along with a power circuit. The power circuit includes an inductor and a capacitor, which serve the purpose of filtering, while an additional inductor is employed for coupling with the DC bus.
	\begin{multline}
		\big[\Delta\dot{X_p}\big] = \big[A_p\big]\big[\Delta X_p\big]+\big[B_{1p}\big]\big[1\big]\big[\Delta  \eta_{c}\big]+\big[K_{pc}\big]\big[B_{1p}\big]\big[1\big]\\ \big[\Delta \zeta_{v}\big]
		+\big[K_{pc}\big]\big[B_{1p}\big]\big[K_{pv}\big]\big[-m_{dc}\big]\big[\Delta P_{dc}\big]
		+\big[K_{pc}\big]\big[B_{1p}\big]\\ \big[-K_{pv}\big]\big[\Delta V_{dc}\big] 
		+\big[K_{pc}\big]\big[B_{1p}\big]\big[H\big]*\big[i_o\big]+\big[B_{1p}\big]\big[-K_{pc}\big]\big[\Delta i_{v}\big]
		\\ +\big[B_{2p}\big]\big[\Delta V_{dc}\big]
	\end{multline}
	where $X_p$ consists three states of the converter as given in eqn \ref{eqn4} and the rest three states considered as follows.
	\begin{equation}\label{eq14}
		\big[\Delta\dot{P_{dc}}\big] = \big[\omega\cdot v_o\big]\big[\Delta i_o\big]+\big[\omega\cdot i_o\big]\big[\Delta v_o\big]-\big[\omega\big]\big[\Delta P_{dc}\big]
	\end{equation}
	\begin{equation}\label{eq15}
		\big[\Delta\dot{\zeta_v}\big] = \big[0\big]\big[\Delta \zeta_{v}\big]+\big[K_{iv}\big]\big[-m_{dc}\big]\big[\Delta P_{dc} \big]+\big[-K_{iv}\big]\big[\Delta V_{dc}\big]
	\end{equation}
	\begin{multline}\label{eq16}
		\big[\Delta\dot{\eta_c}\big] = \big[0\big]\big[\Delta \eta_{c}\big]+\big[K_{ic}\big]\big[1\big]\big[\Delta V_{v}\big]+\big[K_{ic}\big]\big[K_{pv}\big] \\ \big[-m_{dc}\big]\big[\Delta P_{dc}\big] 
		+\big[K_{ic}\big]\big[-K_{pv}\big]\big[\Delta V_{dc}\big]+\big[K_{ic}\big]\big[H\big]*\big[i_o\big] \\ +\big[-K_{ic}\big]\big[\Delta i_{v}\big]
	\end{multline}
	For a whole converter along with controller, the states are as follows: $\big[$ $i_v$,$i_o$,$V_o$,$P_{dc}$,$\zeta_v$ and $\eta_c$  $\big]$
	\section{DC-AC Converter System Modeling for AC Sub-grid}	
	\subsection{State space model of DG-coupled VSC for AC microgrid}
	\subsubsection{State-space model of the power circuitry of VSC} 
\begin{figure}
		\centering
		\includegraphics[width=0.8\linewidth]{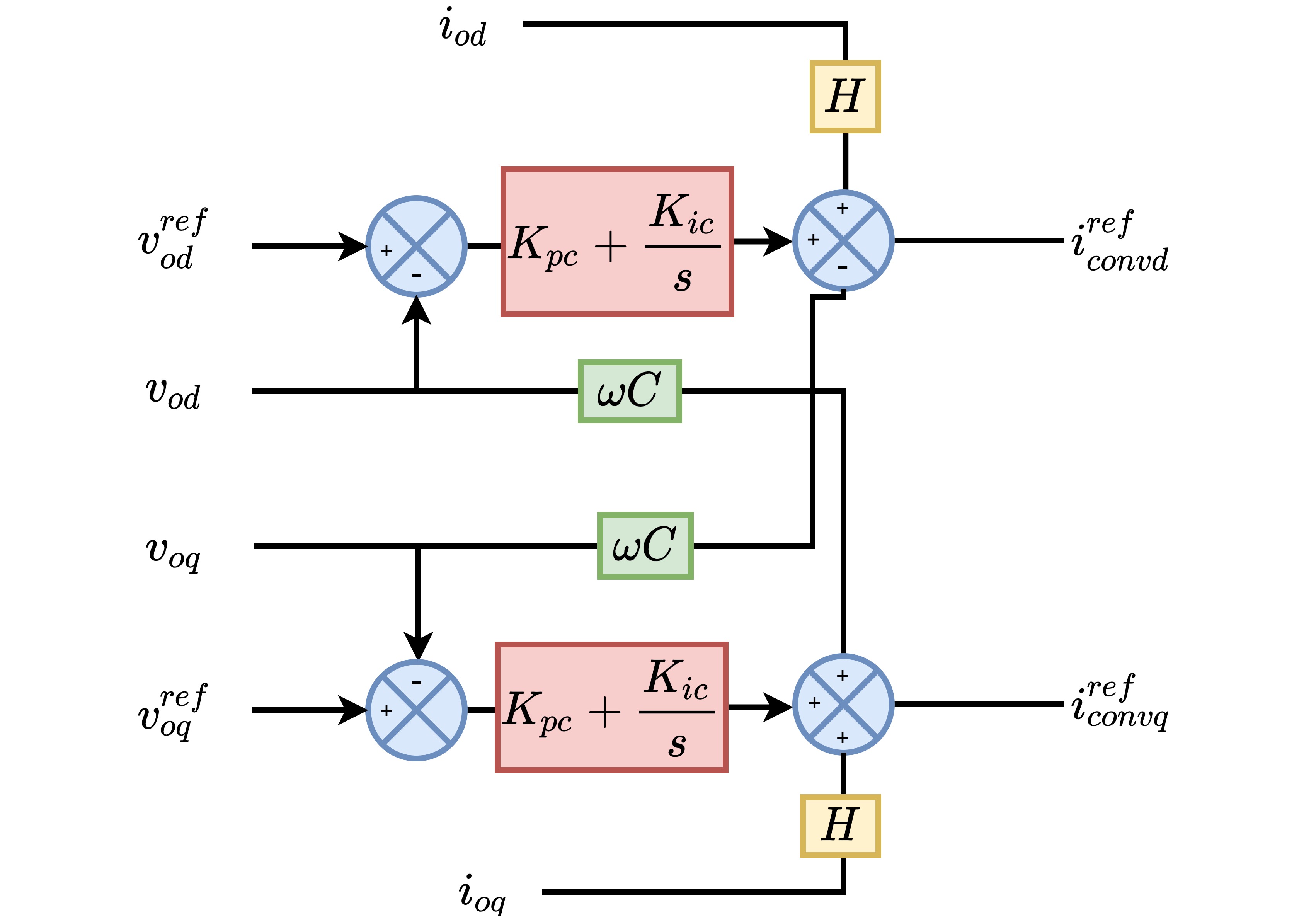}
		\caption{State variable flow for Voltage controller of VSC for AC Sub-grid}
		\label{fig:voltagecontrolleracvsc}
	\end{figure}

	The differential equation for the resistive and inductive voltage drop in the $d$–$q$ reference frame can be obtained by applying Kirchhoff's voltage law (KVL) to the AC side of the VSC. The voltage across the shunt capacitor can be calculated using Kirchhoff's current law (KCL). Consequently, the large-signal model can be represented by the dynamic equations of voltages and currents can be written using Fig. \ref{fig:voltagecontrolleracvsc} as follows:	
	% The equivalent differential equation of the voltage drop across the inductance and resistance in the $d$–$q$ reference frame can be obtained by applying Kirchhoff's voltage law (KVL) to the AC side of the VSC. 	
	\begin{equation}\label{eqn5}
		\frac{\partial i_{inv_{dq}}}{\partial t} = \frac{ v_{inv_{dq}}}{\ L_f} - \frac{ v_{o_{dq}}}{\ L_f} -\frac{R_{f}\cdot i_{inv_{dq}}}{L_f} \pm \omega \cdot i_{inv_{qd}} 
	\end{equation}
	\begin{equation}\label{eqn6}
		\frac{\partial v_{o_{dq}}}{\partial t} = \pm \omega \cdot v_{o_{qd}} + \frac{ i_{inv_{dq}}}{\ C_f} - \frac{ i_{o_{dq}}}{\ C_f} 
	\end{equation}
	\begin{equation}\label{eqn7}
		\frac{\partial i_{o_{dq}}}{\partial t} = \frac{ v_{o_{dq}}}{\ L_o} - \frac{ v_{g_{dq}}}{\ L_o} -\frac{R_{o}\cdot i_{o_{dq}}}{L_o} \pm \omega \cdot i_{o_{qd}} 
	\end{equation}
	where the angular speed ($\omega$) is associated with the $d–q$ reference frame. The AC filter components of VSC are denoted by $R_f$, $L_f$, and $C_f$, while the coupling components of VSC with the PCC are denoted by $R_o$ and $L_o$.
	The following equation shows the small-signal representation of the state-space model in \eqref{eqn5}–\eqref{eqn7} \cite{Pogaku2007}.
	\begin{multline}
		\left[\Delta \dot x_P \right] = \mathcal{A}_{LCL}\left[\Delta x_P \right] + \mathcal{B}_{LCL1}\begin{bmatrix}\Delta v_{inv_d} \\  \Delta v_{inv_q} \end{bmatrix} \\ + \mathcal{B}_{LCL2}\begin{bmatrix}\Delta v_{g_d} \\  \Delta v_{g_q} \end{bmatrix} + \mathcal{B}_{LCL3}\begin{bmatrix} \Delta w \end{bmatrix}
	\end{multline}	
	where 
	$\left[\Delta x_P \right] = \left[\Delta i_{inv_d}\ \Delta i_{inv_q}\ \Delta i_{o_d}\ \Delta i_{o_q}\ \Delta v_{o_d}\ \Delta v_{o_q} \right]^T$\\	
	$\mathcal{A}_{LCL} = \begin{bmatrix}
		\frac{-R_f}{L_f} & w & 0 & 0 & \frac{-1}{L_f} & 0 \\
		-w & \frac{-R_f}{L_f} & 0 & 0 & 0 & \frac{-1}{L_f}\\
		0 & 0 & \frac{-R_o}{L_o} & w & \frac{-1}{L_o} & 0\\
		0 & 0 & -w & \frac{-R_o}{L_o} & 0 & \frac{-1}{L_o}\\
		\frac{1}{C_f} & 0 & \frac{-1}{C_f} & 0 & 0 & w \\
		0 & \frac{1}{C_f} & 0 & \frac{-1}{C_f} & -w & 0 \\
	\end{bmatrix}$;\\
	\\
	$\mathcal{B}_{LCL1} = trans\begin{bmatrix}
		\frac{1}{L_f} & 0 & 0 & 0 & 0 & 0 \\
		0 & \frac{1}{L_f} & 0 & 0 & 0 & 0
	\end{bmatrix}$;\\
	\\
	$\mathcal{B}_{LCL2} = trans\begin{bmatrix}
		\frac{-1}{L_o} & 0 & 0 & 0 & 0 & 0 \\
		0 & \frac{-1}{L_o} & 0 & 0 & 0 & 0
	\end{bmatrix}$;\\	
	$\mathcal{B}_{LCL3} = \left[i_{inv_q}\ -i_{inv_d}\  i_{o_q}\ -i_{o_d}\  v_{o_q}\  -v_{o_d} \right]^T$\\	
	\paragraph{state-space model of voltage controller of VSC} 
	The voltage controller(VC) state-space model can be developed using the preceding modelling procedures. The auxiliary state variables of the integral terms of the VC in the $d-$ and $q-$ directions are repreented by $\Delta\phi_d$ and $\Delta\phi_q$, respectively. In general, to improve system stability, the bandwidth of the loops must increase which is implemented using the feed-forward in voltage controller loops by adding gain blocks \cite{Teodorescu2009}. The linearised state-space model of the voltage controller is as follows:\\
	\begin{equation}
		\begin{bmatrix}  \Delta\dot\phi_d \\  \Delta\dot\phi_q \end{bmatrix}=\left[0\right]\begin{bmatrix}  \Delta\phi_d \\  \Delta\phi_q \end{bmatrix}+\mathcal{B}_{v_1} \begin{bmatrix}  \Delta v_{o_{d}} \\  \Delta v_{o_{q}} \end{bmatrix} + \mathcal{B}_{v_2} \left[ \Delta x_P \right]
	\end{equation}
	\begin{equation}
		\begin{bmatrix}  \Delta\dot i_{o_d} \\  \Delta\dot i_{o_q} \end{bmatrix}=\mathcal{C}_{v}\begin{bmatrix}  \Delta\gamma_d \\  \Delta\gamma_q \end{bmatrix}+\mathcal{D}_{v_1} \begin{bmatrix}  \Delta v_{inv_{d}} \\  \Delta v_{inv_{q}} \end{bmatrix} + \mathcal{D}_{v_2} \left[ \Delta x_P \right]
	\end{equation}
	where \\
	$\mathcal{B}_{v_1} = \begin{bmatrix} K_{iv} & 0 \\ 0 & K_{iv}\end{bmatrix}$;
	$\mathcal{B}_{v_2} = \begin{bmatrix} -1 & 0 & 0 & 0 & 0 & 0 \\ 0 & -1 & 0 & 0 & 0 & 0 \end{bmatrix}$;
	$\mathcal{C}_{v} = \begin{bmatrix} 1 & 0 \\ 0 & 1\end{bmatrix}$;
	$\mathcal{D}_{v_1} = \begin{bmatrix} K_{pv} & 0 \\ 0 & K_{pv}\end{bmatrix}$;\\
	$\mathcal{D}_{v_2} = \begin{bmatrix} 0 & 0 & H & 0 & -K_{pc} & -wC_f\\ 0 & 0 & 0 & H & wC_f & -K_{pc} \end{bmatrix}$\\
	\\
	Below equations represent the state space model of the VSC of the AC microgrid: \\
	\begin{equation}
		\left[\Delta\dot x_{inv} \right] = \mathcal{A}_{inv} \left[\Delta x_{inv} \right] +\mathcal{B}_{inv} \left[\Delta v_{g_{dq}} \right] 
	\end{equation}
	where
	$\Delta v_{g_{dq}} = \begin{bmatrix} v_{g_d} \\ v_{g_q}\end{bmatrix}$	
	\paragraph{State-space representation of the current controller of VSC} 
	The VSC's inner loops which represent current control loops are depicted in \eqref{eq9}. The measured signals $i_{inv_d}$ and $i_{inv_q}$ keep track of the reference values $i_d^{ref}$ and $i_q^{ref}$ through a proportional and integral (PI) controller $(K_{pc} + K_{ic}/s)$. The current controller loops' state-space model includes two additional state variables, $\Delta\gamma_ d$ and $\Delta\gamma_q$, to reflect the current controller's integral term in the $d-$ and $q-$ directions, respectively	
	\begin{equation}\label{eq9}
		L_f\frac{\partial i_{inv_{dq}}}{\partial t} = \begin{pmatrix} K_{pc} + \frac{K_{ic}}{s} \end{pmatrix} \times \begin{pmatrix} i_{dq}^{ref} - i_{inv_{dq}} \end{pmatrix}
	\end{equation}
	The linearised state-space model for the current controller can be expressed as follows:
	\begin{equation}
		\begin{bmatrix}  \Delta\dot\gamma_d \\  \Delta\dot\gamma_q \end{bmatrix}=\left[0\right]\begin{bmatrix}  \Delta\gamma_d \\  \Delta\gamma_q \end{bmatrix}+\mathcal{B}_{c_1} \begin{bmatrix}  \Delta v_{inv_{d}} \\  \Delta v_{inv_{q}} \end{bmatrix} + \mathcal{B}_{c_2} \left[ \Delta x_P \right]
	\end{equation}
	\begin{equation}
		\begin{bmatrix}  \Delta\dot v_{inv_d} \\  \Delta\dot v_{inv_q} \end{bmatrix}=\mathcal{C}_{c}\begin{bmatrix}  \Delta\gamma_d \\  \Delta\gamma_q \end{bmatrix}+\mathcal{D}_{c_1} \begin{bmatrix}  \Delta v_{inv_{d}} \\  \Delta v_{inv_{q}} \end{bmatrix} + \mathcal{D}_{c_2} \left[ \Delta x_P \right]
	\end{equation}
	where 
	$\mathcal{B}_{c_1} = \begin{bmatrix} K_{ic} & 0 \\ 0 & K_{ic}\end{bmatrix}$;\\
	$\mathcal{B}_{c_2} = \begin{bmatrix} -K_{ic} & 0 & 0 & 0 & 0 & 0 \\ 0 & -K_{ic} & 0 & 0 & 0 & 0 \end{bmatrix}$;
	$\mathcal{C}_{c} = \begin{bmatrix} 1 & 0 \\ 0 & 1\end{bmatrix}$;\\
	$\mathcal{D}_{c_1} = \begin{bmatrix} K_{pc} & 0 \\ 0 & K_{pc}\end{bmatrix}$\\ \ \ \&
	$\mathcal{D}_{c_2} = \begin{bmatrix} -K_{pc} & -wL_f & 0 & 0 & 1 & 0 \\ wL_f & -K_{pc} & 0 & 0 & 0 & 1 \end{bmatrix}$
	\begin{figure}
		\centering
		\includegraphics[width=0.8\linewidth]{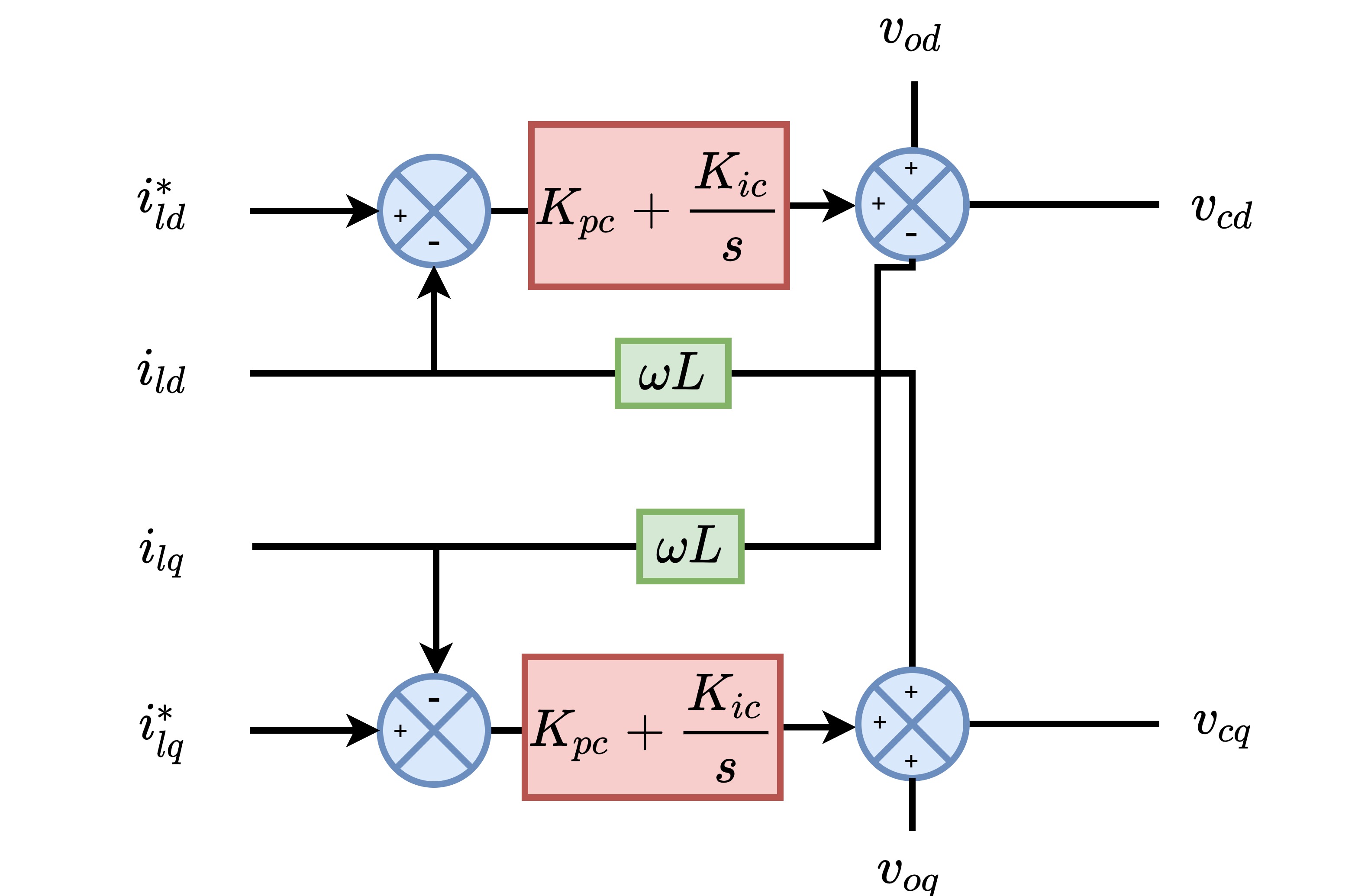}
		\caption{State variable flow for Current controller of VSC for AC Sub-grid}
		\label{fig:currentcontrolleracvsc}
	\end{figure}
	\paragraph{State-space model of power controller for VSC} 
	The power controller is made up of of three integrators which can be represented in differential equations in mathematical form as follows.
	\begin{equation}
		\dot P = \begin{pmatrix} \frac{3}{2} \end{pmatrix}\left(w_f \left( v_{o_d}\cdot i_{o_d}+  v_{o_q} \cdot i_{o_q}- P\right)\right)
	\end{equation}
	\begin{equation}
		\dot Q = \begin{pmatrix} \frac{3}{2} \end{pmatrix}\left(w_f \left( v_{o_d}\cdot i_{o_q}-  v_{o_d} \cdot i_{o_q}- Q\right)\right)
	\end{equation}	
	Using the differential equation representation of the power controller, state-space model can be written as follows:\\
	\begin{equation}
		\begin{bmatrix} \Delta\dot\theta \\\Delta\dot P\\\Delta\dot Q \end{bmatrix} = \mathcal{A}_p\begin{bmatrix} \Delta\theta \\\Delta P\\\Delta Q \end{bmatrix} + \mathcal{B}_{p_1} \cdot\left[\Delta x_p \right]    
	\end{equation}
	The power controller's output equations are associated with the PCC voltages of the system which is shown in these below equations:
	\begin{equation}
		\begin{bmatrix} \Delta v_{o_d} \\ \Delta v_{o_q} \end{bmatrix} = \mathcal{C}_p\begin{bmatrix} \Delta Q \\\Delta P\end{bmatrix}    
	\end{equation}
	where 
	$\mathcal{A}_p = \begin{bmatrix} -w_f & 0 & 0 \\ 0 & -w_f & 0 \\ -n_p & 0 & 0 \end{bmatrix}$;
	$\mathcal{C}_p = \begin{bmatrix} 0 & -n_p & 0 \\ 0 & 0 & 0 \end{bmatrix}$; \\
	$\mathcal{B}_{p_1} = \left(\frac{3}{2}\right)\begin{bmatrix} 0 & 0 & w \cdot v_{o_d} & w \cdot v_{o_q} & w \cdot i_{o_d} & w \cdot i_{o_q} \\ 0 & 0 & -w \cdot v_{o_q} & w \cdot v_{o_d} & w \cdot i_{o_q} & -w \cdot i_{o_d} \\ 0 & 0 & 0 & 0 & 0 & 0 \end{bmatrix}$\\
	\subsection{State-space model of the Interlinking Converter}
\begin{figure}
		\centering
		\includegraphics[width=1.0\linewidth]{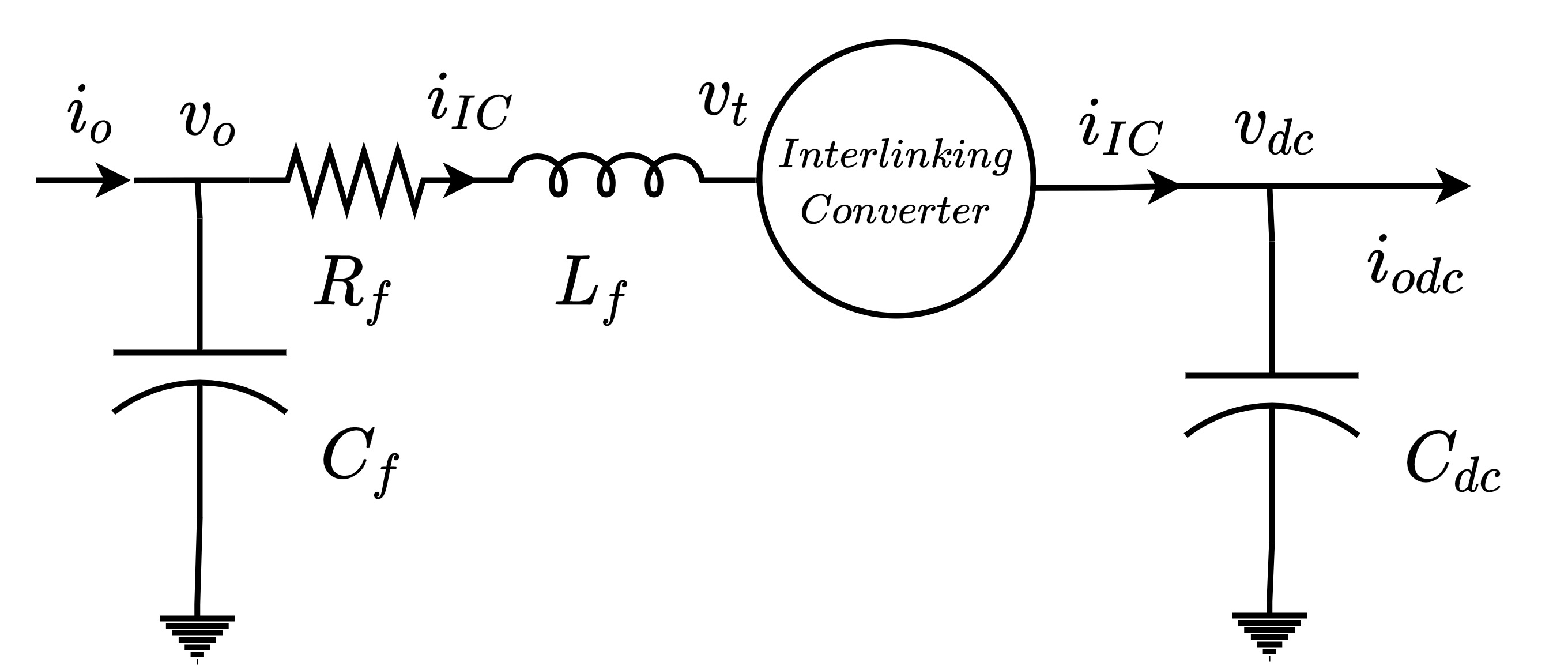}
		\caption{Interlinking converter equivalent diagram}
		\label{fig:icmodel}
	\end{figure}

	Except for the power circuit of the state-space model, the Interlinking converter (IC) model is analogous to the model of the voltage source converter in the AC microgrid. \textcolor{black}{The power control for the IC, as shown in Fig. \ref{fig:icmodel}, is established as per swing equation emulating the behavior of synchronous machine, shown by the equations below.}
	\begin{equation}
		\frac{\partial w}{\partial t} = \frac{P_{ref}}{2J}-\frac{P_{IC}}{2J}-\frac{K_d\cdot w_{vsm}}{2J}+\frac{K_d\cdot w_{g}^{*}}{2J}
	\end{equation}
	\begin{equation}
		\frac{\partial \delta}{\partial t} =  w_{vsm}
	\end{equation}
	\subsubsection{State space model of the power circuit of the IC}
	The large signal model of the Interlinking Converter (IC) can be represented by the dynamic equation of the voltage and current of IC and can be written as follows:
	\begin{equation}
		\frac{\partial i_{IC_{dq}}}{\partial t} = \frac{V_{odq}}{L_f}-\frac{m_{dq}\cdot V_{dc}}{L_f}-\frac{R_f\cdot i_{IC_{dq}}}{L_f}\pm w\cdot i_{IC_{dq}}
	\end{equation}
	\begin{equation}
		\frac{\partial v_{IC_{dq}}}{\partial t} = \pm w \cdot V_{qd}+\frac{i_{IC_{dq}}}{C_f}-\frac{i_{o_{dq}}}{C_f}
	\end{equation}
	\begin{equation}
		C_{dc}\frac{\partial v_{dc}}{\partial t} = 1.5m_{d}{i_{IC_d}}-1.5m_{q}{i_{IC_q}}-i_{o_{dc}}
	\end{equation}
	The interlinking converter power circuit's small-signal state-space representation is as follows::
	\begin{equation}
		\begin{aligned}
			\left[\Delta\dot x_{p_{IC}} \right] = \mathcal{A}_{p_{IC}} \left[\Delta x_{p_{IC}} \right] +\mathcal{B}_{p_{IC1}} \begin{bmatrix}  \Delta m_{d} \\ \Delta m_{q}  \end{bmatrix} +\mathcal{B}_{p_{IC2}} \begin{bmatrix}  \Delta i_{o_d} \\ \Delta i_{o_q}  \end{bmatrix}\\ + \mathcal{B}_{p_{IC3}} \left[\Delta w\right] + \mathcal{B}_{p_{IC4}} \left[\Delta i_{o_{dc}}\right]
		\end{aligned}
	\end{equation}
	where
	$\left[\Delta x_{p_{IC}} \right] = \left[ \Delta i_{ic_d}\ \Delta i_{ic_q}\ \Delta v_{ic_d}\ \Delta v_{ic_q}\ \Delta v_{dc} \right]$\\ \\
	$\mathcal{A}_{p_{IC}} = \begin{bmatrix} \frac{-R_f}{L_f} & w & \frac{-1}{L_f} & 0 & \frac{m_d}{L_f}\\
		-w & \frac{-R_f}{L_f} & 0 & \frac{-1}{L_f} & \frac{m_q}{L_f}\\
		\frac{1}{C_f} & 0 & 0 & w & 0\\
		0 & \frac{1}{C_f} & -w & 0 & 0\\
		\frac{1}{C_f} & 0 & \frac{-1}{C_f} & 0 & 0\\
		\frac{-1.5m_d}{C_{dc}} & \frac{-1.5m_q}{C_{dc}} & 0 & 0 & 0 \end{bmatrix}$\\ \\
	$\mathcal{B}_{p_{IC1}} = \begin{bmatrix} \frac{V_{dc}}{L_f} & 0 & 0 & 0 & \frac{-1.5 i_{od}}{C_{dc}}\\ 
		0 & \frac{V_{dc}}{L_f} & 0 & 0 & \frac{-1.5 i_{oq}}{C_{dc}}\end{bmatrix}^T$\\ \\
	$\mathcal{B}_{p_{IC2}} = \begin{bmatrix} 0 & 0 & \frac{-1}{C_f} & 0 & 0\\
		0 & 0 & 0 & \frac{-1}{C_f} & 0\\ \end{bmatrix}^T$\\ \\
	$\mathcal{B}_{p_{IC3}} = \begin{bmatrix} i_{ic_q} &  -i_{ic_d} & V_q & -V_d & 0\\\end{bmatrix}^T$\\ \\
	$\mathcal{B}_{p_{IC4}} = \begin{bmatrix} 0 &  0 & 0 & 0 & \frac{1}{C_{dc}} \end{bmatrix}^T$\\
	\subsubsection{State-space modelling of the controller of Interlinking Converter}
	The procedures to obtain the IC controller's state-space modelling can be formulated as explained in \cite{yan2023} and not explored in this paper to avoid redundancy and space constraints. The overall states of the controller of the IC, are as shown, are made up of state variables based on the equation shown below.
	\begin{equation}
		\begin{aligned}
			\left[\Delta\dot x_{p_{IC}} \right] = \mathcal{A}_{inv_{IC}} \left[\Delta x_{p_{IC}} \right] +\mathcal{B}_{inv_{IC}} \begin{bmatrix}  \Delta io_{d} \\ \Delta io_{q} \\ \Delta io_{dc} \end{bmatrix}
		\end{aligned}
	\end{equation}	
	\section{Hybrid Microgrid Test System}
\begin{figure}[H]
		\centering
		\includegraphics[width=0.9\linewidth]{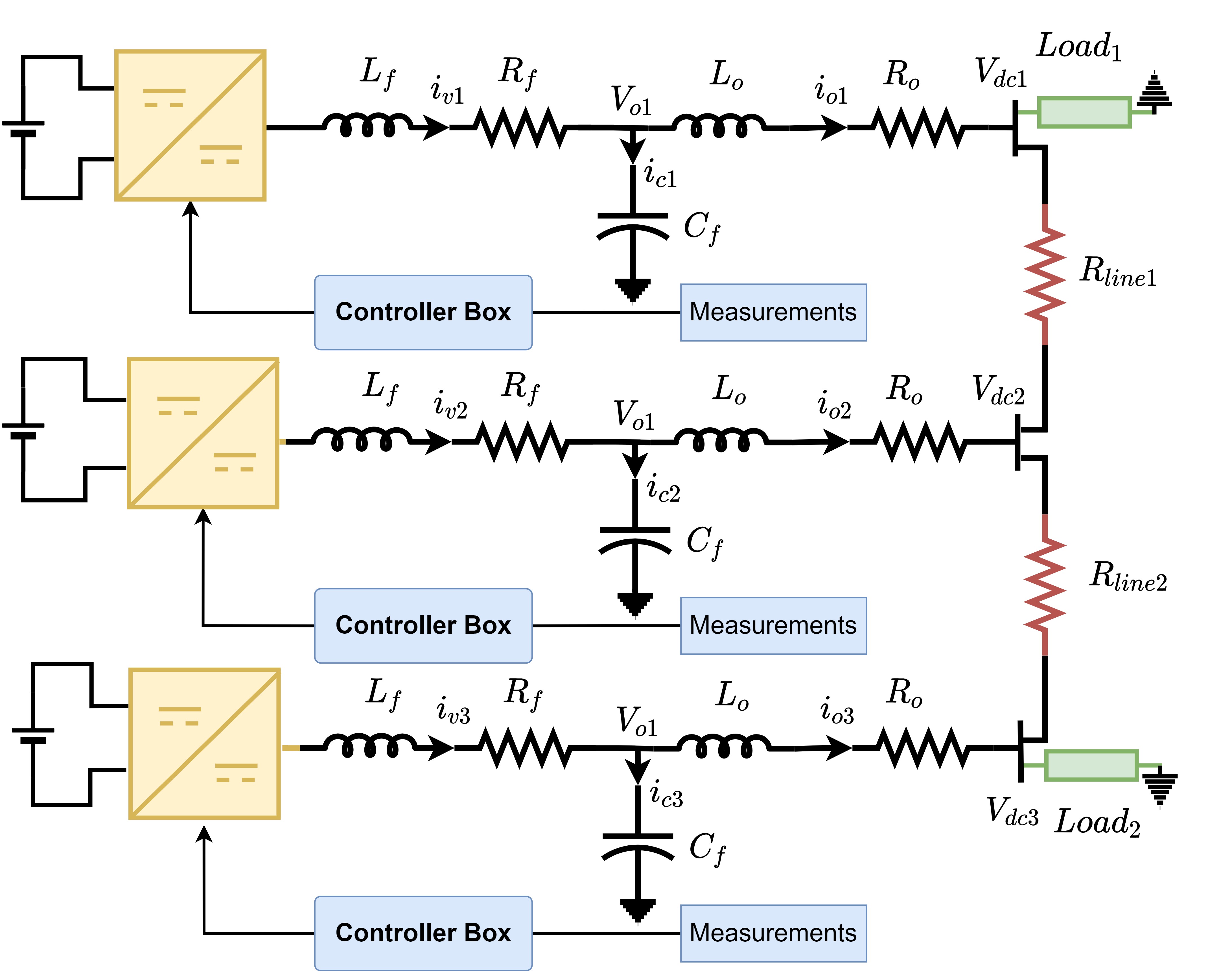}
		\caption{An equivalent representation of DC sub-grid using three DC-DC Converters as modelled}
		\label{fig:convertermodelcircuit}
	\end{figure}

\begin{figure}[H]
		\centering
		\includegraphics[width=0.9\linewidth]{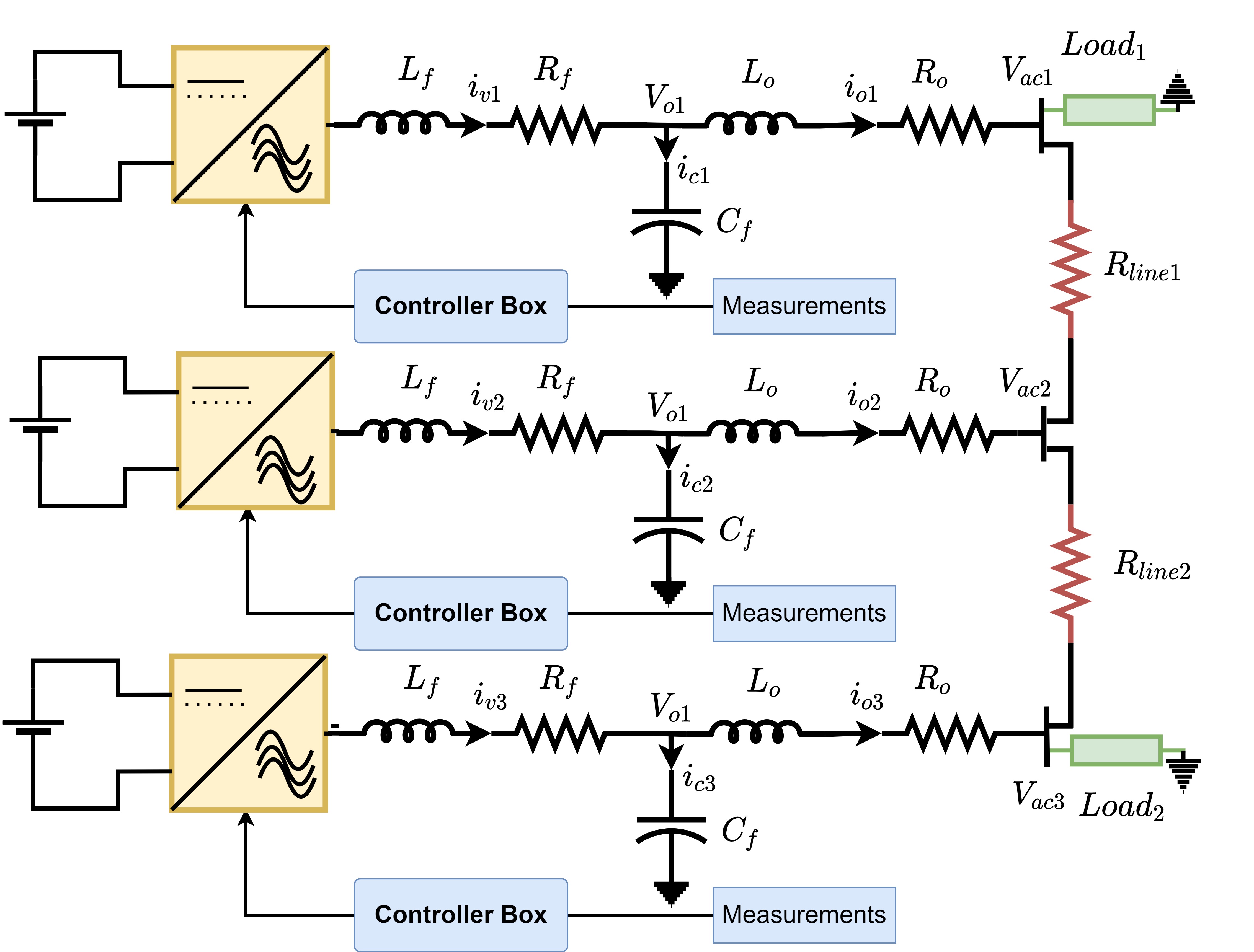}
		\caption{An equivalent representation of AC sub-grid using three DC-AC Converters as modelled}
		\label{fig:convertermodelcircuitacside}
	\end{figure}

	The hybrid microgrid test system consisting of DC sub-grid and AC sub-grid with interlinking converter for bidirectional power flow in between the sub-grids. The DC sub-grid contains three DC-DC converter and two loads as shown in Fig. \ref{fig:convertermodelcircuit}, are established in MATLAB environment using the state-space model. Each DC converter is connected to filter circuitry for better voltage and current waveforms. Two resistive lines interconnect the DC buses. The AC sub-grid comprises three DC-AC converters and two loads situated at buses 1 and 3, as depicted in Fig. \ref{fig:convertermodelcircuitacside}, modeled in MATLAB. Each converter is associated with filter circuitry and connected to its respective bus. Bus 2 of the AC sub-grid serves as the connection point for the interlinking converter, which facilitates bidirectional power flow between bus 2 of the DC sub-grid and the AC sub-grid.
	The filter, lines, load, and controller parameters are represented as shown in table \ref{tab:filter}, \ref{tab:line}, and \ref{tab:controller}, respectively.	
	\begin{table}[H]
		\captionof{table}{Filter and Connector $RLC$ Parameters} \label{tab:filter} 
		\begin{center}
			\begin{tabular}{|c|c|c|c|c|c|}
				\toprule[1.5pt]
				Parameters & $R_f$ & $L_f$ & $C_f$ & $R_o$ & $L_o$\\
				\midrule
				Values  & 0.01 $\Omega$ & 0.03 $H$ & 20e-6 $\mu F$	& 0.01 $\Omega$ & 0.05 $H$ \\
				\bottomrule[1.5pt]
			\end{tabular}
		\end{center}
	\end{table}
	\begin{table}[H]
		\captionof{table}{Line and Load Parameters} \label{tab:line} 
		\begin{center}
			\begin{tabular}{|c|c|c|c|c|}
				\toprule[1.5pt]	
				Parameters & $R_{line1}$ & $L_{line1}$ & $R_{load1}$ & $R_{load2}$\\
				\midrule
				Values  & 0.04 & 0.01 & 20	 & 50\\
				\bottomrule[1.5pt]
			\end{tabular}
		\end{center}
	\end{table}
	\begin{table}[H]
		\captionof{table}{Controller Parameters} \label{tab:controller} 
		\begin{center}
			\begin{tabular}{|c|c|c|c|c|c|c|}
				\toprule[1.5pt]	
				Parameters & $K_{pv}$ & $K_{iv}$ & $K_{pc}$ & $K_{ic}$ & $\omega$ & $m_{dc}$ \\
				\midrule
				Values & 1 & 33 & 0.2 & 120 & 2*$\pi$*10 & 1e-3 \\
				\bottomrule[1.5pt]
			\end{tabular}
		\end{center}
	\end{table}
	\section{Pole-Placement using state Feedback}
\begin{figure}[H]
		\centering
		\includegraphics[width=0.9\linewidth]{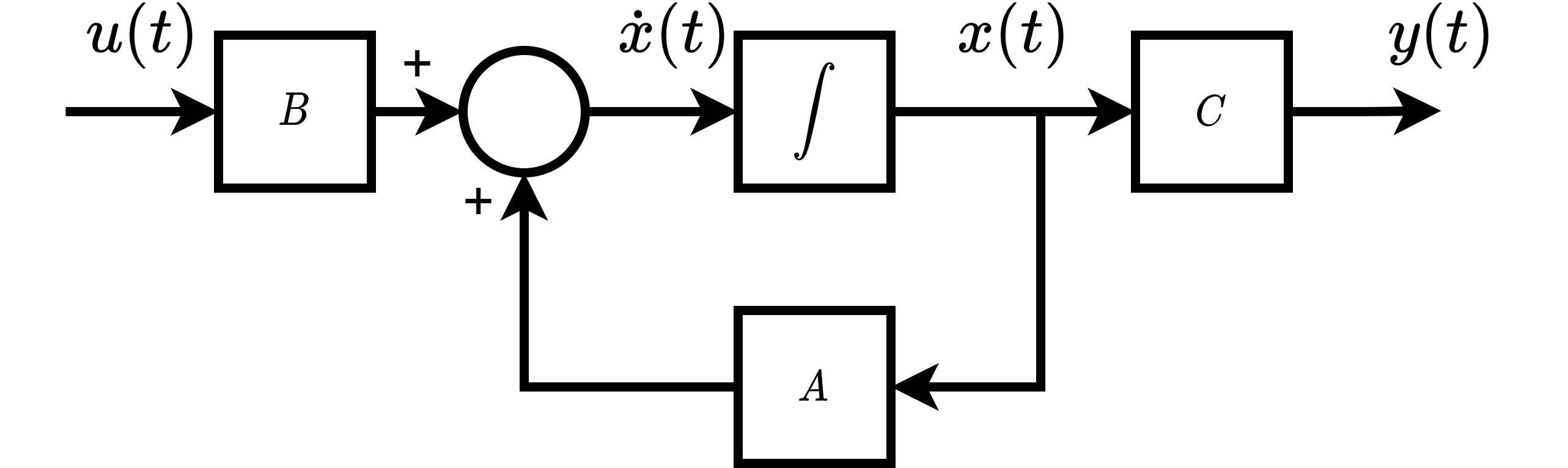}
		\caption{State variable form of a control system}
		\label{fig:statevariable}
	\end{figure}

	State-feedback pole placement is frequently referred to as the inverse eigenvalue problem in control theory. In practical control system design, it is essential to ensure that the closed-loop pole locations remain relatively insensitive to perturbations in the system's coefficient matrices. This requirement inherently limits the available degrees of freedom in the pole-placement process.
    Consider a linear time-invariant (LTI) multi-input multi-output (MIMO) system described by the following state-space representation:
	\begin{equation*}
		\mathcal{P}{x} = Ax+Bu
	\end{equation*}
	\begin{equation*}
		y = Cx+Du
	\end{equation*}
	where $x$, $u$ are $n$ and $m$ dimensional vectors respectively. The state variable structure is shown in Fig. \ref{fig:statevariable}.
	$\mathcal{P}$ signifies the differential operator, commonly denoted by $\frac{d}{dt}$, in the context of continuous-time systems, or the shift (delay) operator, typically denoted by $z^{-1}$, in the context of discrete-time systems. The dynamic behavior of the system is governed by the location of its poles, which correspond to the eigenvalues of the system matrix $A$.
	In a state-space representation of a system, it is commonly assumed that there is no direct feedthrough from the input to the output, i.e., the feedthrough matrix $D=0$. The placement of the system's eigenvalues, or poles, at specific desired locations facilitates the attainment of properties such as improved stability. This objective can be realized through state feedback control, wherein the input is modified accordingly \cite{Kautsky1985}
	\begin{equation}
		u = \mathcal{F}x+v
	\end{equation}
	where $\mathcal{F}$ is the feedback or gain matrix to modify the system dynamics described as 
	\begin{equation}
		\mathcal{P}{x} = (A+B\mathcal{F})x+Bv
	\end{equation}
	where with input $v$ we obtain desired pole locations
	\begin{figure}[H]
		\centering
		\includegraphics[width=0.9\linewidth]{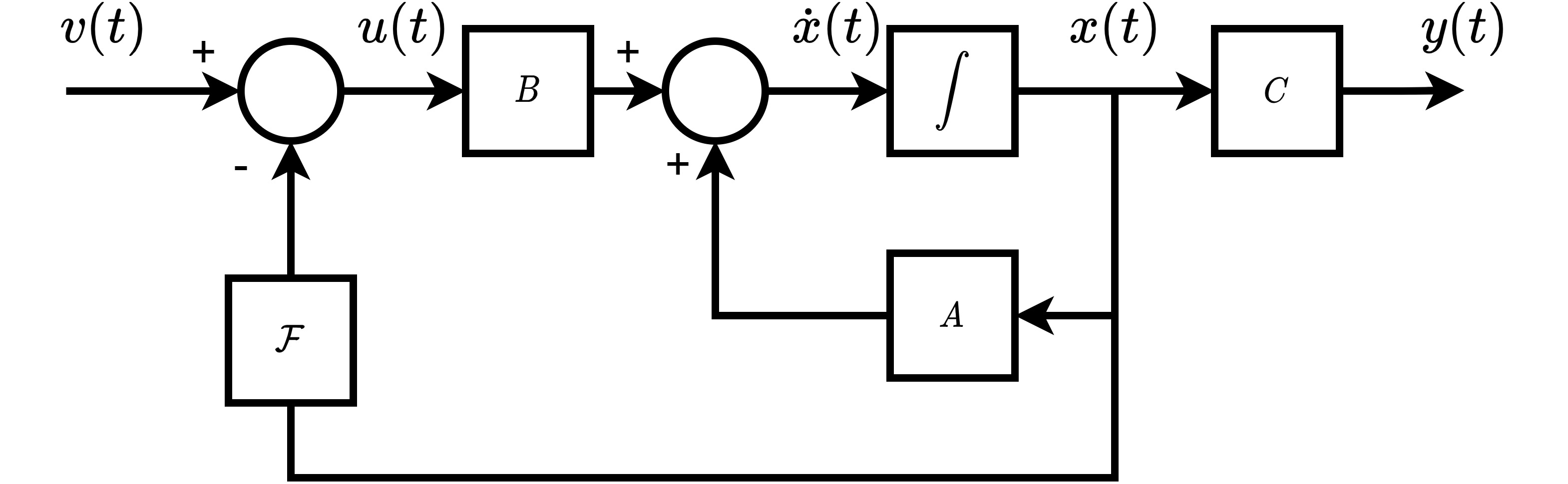}
		\caption{State feedback structure of a control system}
		\label{fig:statefeedback}
	\end{figure}
	The state feedback structure is shown in Fig. \ref{fig:statefeedback}, which demonstrates the closed-loop pole placement by varying the $\mathcal{F}$ matrix as per desired closed-loop pole locations.\\
	\begin{Theorem}
		Given $\Lambda = diag(\lambda_1, \lambda_2, ... . \lambda_n)$ and $X$ is non-singular there exists $\mathcal{F}$, a solution if 
		$U_1^T(AX-X\Lambda)=0$
		\\
		where $B$ = $\big[U_0\ U_1\big]$$\big[Z\ 0\big]^T$ with $U$ orthogonal and $Z$ non-singular, then $\mathcal{F}$ is explicitly given by $\mathcal{F} = Z^{-1}U_0^T(X\Lambda X^{-1}-A)$.
	\end{Theorem}	
	\section{Results and Discussions}
	To facilitate the dynamic analysis of the converter-based DC microgrid, the state-space model formulated in Section II is implemented within the MATLAB environment. The parameters utilized for modeling the test system are detailed in Tables I, II, and III in Section III. A state-space representation of the converter is first developed; subsequently, these converter models are integrated to construct the overall test microgrid. The model includes two transmission lines connecting buses 1–2 and 2–3, as well as two loads connected at buses 1 and 3, respectively. In addition, a simulation-based model is developed using the same set of parameters as the state-space model, in order to evaluate the regulation and tracking performance of the system.
	\subsection{Eigenvalue Analysis} 
\begin{figure}
		\centering
		\begin{minipage}{0.45\textwidth}
			\centering
			\includegraphics[width=1.0\linewidth]{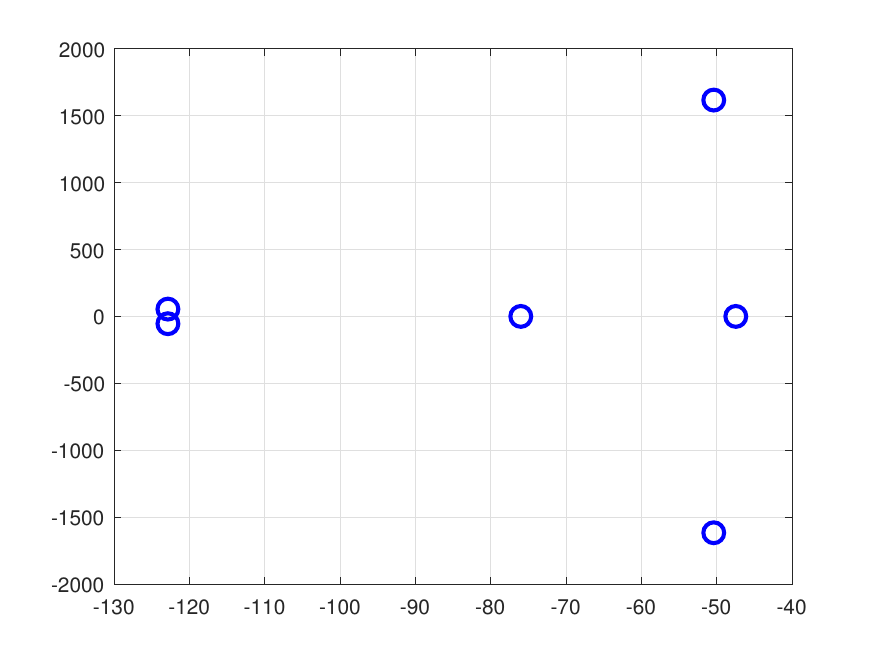}
			\caption{Eigenvalues of the DC-DC converter, six eigenvalues corresponds to six states}
			\label{fig:conveigenvalues}
		\end{minipage}
		\begin{minipage}{0.45\textwidth}
			\centering
			\includegraphics[width=1.0\linewidth]{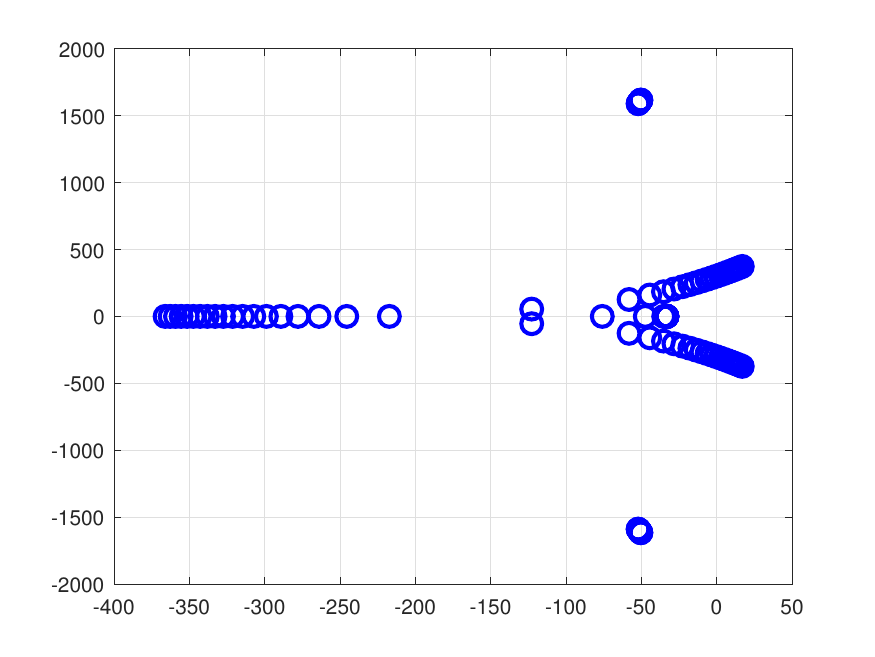}
			\caption{Eigenvalues of the DC-DC converter with varying droop}
			\label{fig:conveigenvalueswithvaryingdroop}
		\end{minipage}
	\end{figure}

\begin{figure}
		\centering
		\includegraphics[width=1.0\linewidth]{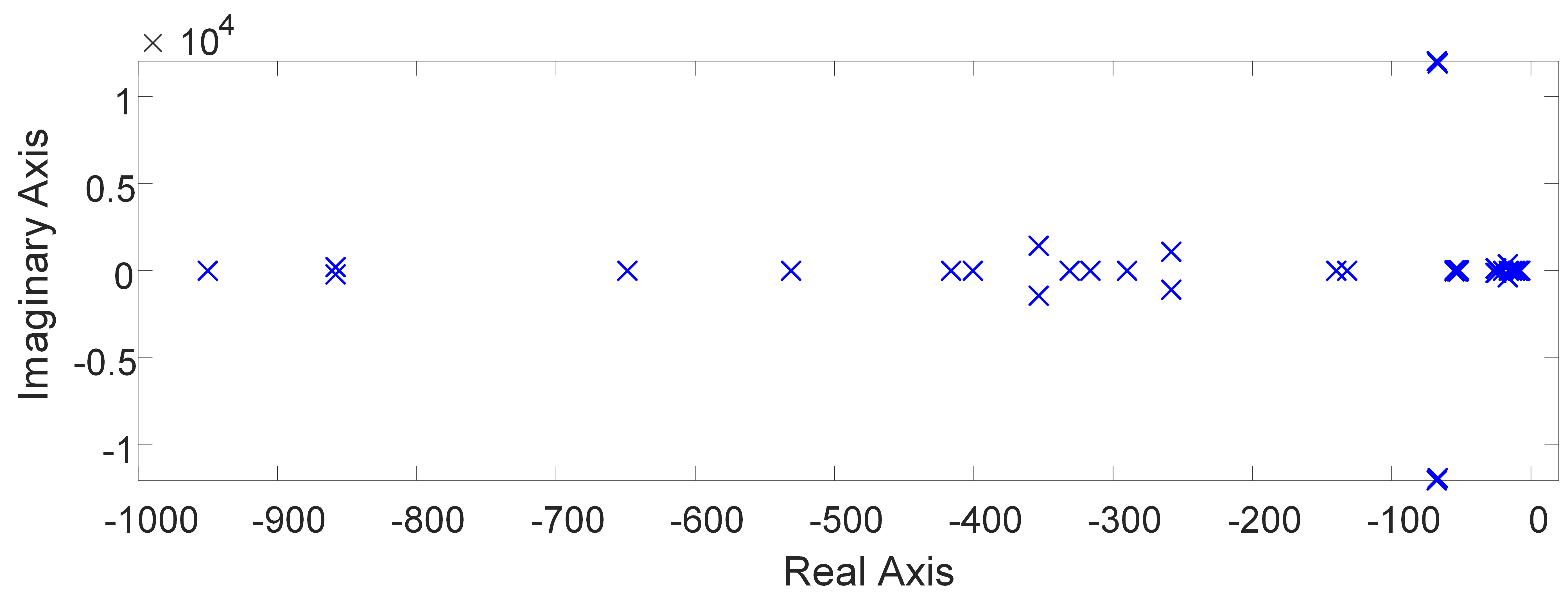}
		\caption{Eigenvalues of the hybrid microgrid test system}
		\label{fig:drawing1}
	\end{figure}

	The developed model of the DC, Hybrid microgrid is analyzed for stability by studying its eigenvalues as shown in Fig. \ref{fig:conveigenvalues},\ref{fig:drawing1}. The system state matrix is further analysed with QR decomposition and shifted QR decomposition to find the eigenvalues and eigenvectors. Using the QR decomposition, the state matrix is reduced to a Hessenberg matrix having the same eigenvalues making the R matrix upper triangular matrix from which the eigenvalues can be read from its diagonal elements. For the inherent algorithms can be referred in \cite{crow2002computational}.
	The circles represent the eigenvalues of the converter model operating in a closed-loop configuration. The relationship between the power droop gain and the corresponding eigenvalues is illustrated in Fig. \ref{fig:conveigenvalueswithvaryingdroop}. Preliminary analysis reveals that, as the droop gain varies, a pair of complex conjugate poles initially located in the left half of the complex plane cross over into the right half, indicating a transition to instability. This phenomenon is characteristic of a Hopf bifurcation. For the system to maintain steady-state operation, it is essential that all eigenvalues remain in the left half of the complex plane, corresponding to well-damped dynamic modes \cite{gopal2002control}. Adequate damping of all modes ensures the suppression of oscillations arising from state perturbations, thereby promoting stable system behavior.
	The variation of eigenvalues with respect to the droop coefficient is plotted. The droop coefficient varies from 1e-3 to 10e-1 with a step length of 5e-2. From this analysis, an eigenvalue-centered stability region with respect to the droop coefficient can be formulated.
	The overall hybrid microgrid test system eigenvalue analysis is shown in Fig. \ref{fig:drawing1}.
	\subsection{Pole-Zero Analysis}	
\begin{figure}[H]
		\begin{minipage}{0.45\textwidth}
			\centering
			\includegraphics[width=1.0\linewidth]{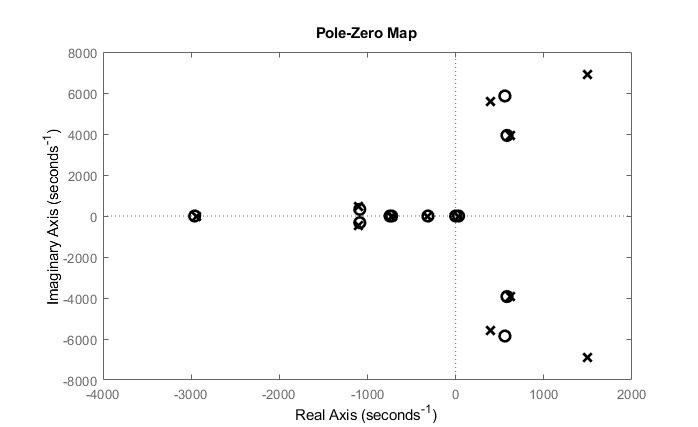}
			\caption{Pole-Zero map without state feedback controller}
			\label{fig:pzmapopen}
		\end{minipage}
		\begin{minipage}{0.45\textwidth}
			\centering
			\includegraphics[width=1.0\linewidth]{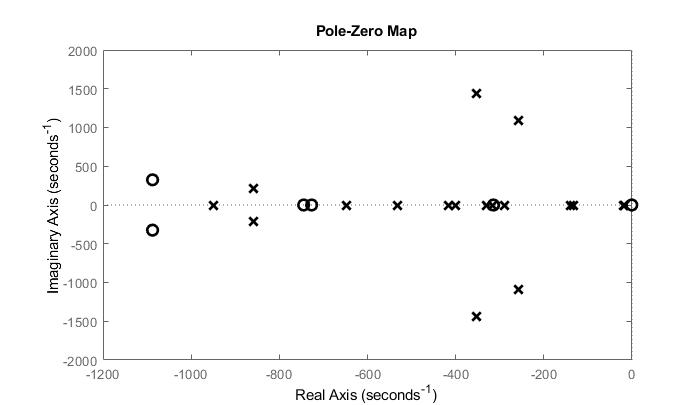}
			\caption{Pole-Zero map with state feedback controller}
			\label{fig:pzmapclosed}
		\end{minipage}
	\end{figure}

	For the purpose of stability analysis, the pole-zero map of the open-loop system is presented in Fig. \ref{fig:pzmapopen}. Following the implementation of state feedback control, the pole-zero configuration is re-evaluated. It is observed that all poles and zeros are relocated to the left half of the complex plane, indicating a stable system response. The corresponding pole-zero map of the closed-loop system is depicted in Fig. \ref{fig:pzmapclosed}.

	The $\mathcal{F}$ matrix used to place the poles in desired locations is given as below
	$\big[\mathcal{F} = 0.91004\        1.526\    0.0019204\    0.0011956\      -1.7949\     -0.18273\   4.8295exp+08\   8.0175exp+08\   9.9502exp+06\   1.8861exp+05\	-6.6886exp+08\   3.7264exp+08\     -0.50278\    -0.84502\   -0.0012685\  -0.00047027\      0.71507\     0.056547\	 \big]$.
	After the pole placement, the pole-zero map is plotted in Fig. \ref{fig:pzmapclosed}, and it can be observed that all the poles lie in the left half plane.
	\subsection{Step response of the system states using state feedback}
	The step response of a system provides valuable insight into the dynamic behavior of system states in response to a unit-magnitude step input. In the open-loop configuration, due to the interaction between line dynamics and converter states, the system exhibits instability. This is evidenced by negatively damped oscillations observed in the step response under non-zero initial conditions. By applying state feedback control, the system poles are reallocated to the left half of the complex plane with an accuracy of 10\% relative to their assigned locations, thereby stabilizing the system dynamics.
    The open-loop system can be described by the transfer function $G(s)$ where the step response is the inverse Laplace transform of $\frac{G(S)}{s}$. Given the system's instability, state feedback is applied to place the poles of the closed-loop system in the left half of the complex plane, ensuring stability. The pole placement is achieved with an accuracy of 10\% relative to the assigned pole locations, based on the desired damping ratio 
$\zeta$and natural frequency $\omega_n$, as specified by the control design.

The closed-loop system’s transfer function can be represented as $G_{cl}(s)=\frac{G(s)}{1+G(s)H(s)}$ where
$H(s)$ is the feedback path transfer function. The step response of the closed-loop system with state feedback results in a stable, well-damped response, as the poles are now located in the left half-plane with the desired damping ratio, 
$\zeta\geq 0.7$, which indicates a critically damped or over-damped response, ensuring the absence of oscillations.
	\section{Conclusion}
	To plan the infrastructure of a hybrid microgrid, the mathematical models of components like converters, loads, and interconnects were investigated. After analysis, a model for the hybrid microgrid was proposed, and small-signal stability analysis was conducted. Without a controller, the models are usually unstable. The models of the aforementioned components, along with their controllers, were analyzed to study the stability of the system. The different controllers were explained using a matrix approach and also using differential equations. A state-space model for the considered converter was developed and integrated into the load and line parameters to form a complete hybrid microgrid system. Further, an eigenvalue study without the controller was conducted, and a pole-placement using a state feedback controller was designed to make the test system Lyapunov stable, including line and load dynamics. The controller designs the closed-loop poles in such a way that the eigenvalues lie in the left-half plane. To obtain a well-defined time behavior from the hybrid microgrid system, a step response for the system states was observed for performance evaluation.

    \bibliographystyle{ieeetr}
    \bibliography{library}
\end{document}